# Advancing Software Security and Reliability in Cloud Platforms through AI-based Anomaly Detection

Sabbir M. Saleh
Computer Science
University of Western Ontario
London, ON, Canada
ssaleh47@uwo.ca

Ibrahim Mohammed Sayem
Computer Science
University of Western Ontario
London, ON, Canada
isayem@uwo.ca

Nazim Madhavji
Computer Science
University of Western Ontario
London, ON, Canada
madhavji@gmail.com

John Steinbacher
IBM Canada Lab
Toronto, ON, Canada
jstein@ca.ibm.com

## ABSTRACT

Continuous Integration/Continuous Deployment (CI/CD) is fundamental for advanced software development, supporting faster and more efficient delivery of code changes into cloud environments. However, security issues in the CI/CD pipeline remain challenging, and incidents (e.g., DDoS, Bot, Log4j, etc.) are happening over the cloud environments. While plenty of literature discusses static security testing and CI/CD practices, only a few deal with network traffic pattern analysis to detect different cyberattacks. This research aims to enhance CI/CD pipeline security by implementing anomaly detection through AI (Artificial Intelligence) support. The goal is to identify unusual behaviour or variations from network traffic patterns in pipeline and cloud platforms. The system shall integrate into the workflow to continuously monitor pipeline activities and cloud infrastructure. Additionally, it aims to explore adaptive response mechanisms to mitigate the detected anomalies or security threats. This research employed two popular network traffic datasets, CSE-CIC-IDS2018 and CSE-CIC-IDS2017. We implemented a combination of Convolution Neural Network (CNN) and Long Short-Term Memory (LSTM) to detect unusual traffic patterns. We achieved an accuracy of 98.69% and 98.30% and generated log files in different CI/CD pipeline stages that resemble the network anomalies affected to address security challenges in modern DevOps practices, contributing to advancing software security and reliability.

## CCS CONCEPTS

• Computing methodologies~ Machine learning~ Machine learning approaches~ Neural networks

## 1 Introduction

In today's fast-paced software development world, the CI/CD pipeline becomes the foundation of delivering applications quickly and efficiently. This integration, often guided by DevOps principles, streamlines collaboration between development and operations teams, allowing for rapid deployment of code changes [1, 2]. Cloud computing has revolutionised the way organisations manage their IT infrastructure. Many businesses, healthcare providers, and government agencies rely on cloud platforms to host critical applications and data. However, new cybersecurity risks come with the benefits of cloud computing [3]. Despite the numerous advantages of the CI/CD pipeline, it also presents significant security challenges [4], especially when deployed in cloud environments. Threats like ransomware attacks, denial of service (DoS), distributed denial of service (DDoS), Bot, Cross-Site Scripting (XSS), and supply chain vulnerabilities have become increasingly common, posing severe challenges to the security of cloud environments.

Cloud platforms face persistent security threats, including downgrade attacks on Transport Layer Security (TLS), a crypto protocol [4] (e.g., ROBOT, DROWN, POODLE) and software supply chain attacks [5] (e.g., Log4j, SolarWinds, CodeCov, etc.). These breaches can go undetected for extended periods, leading to significant consequences. For example, the CodeCov attackers exploited a configuration flaw, gaining unauthorised access to the source code repositories of 23,000 clients [6].

In 2021, Verizon Communications and Facebook experienced cloud-related security incidents that exposed user data due to Amazon Web Service (AWS) vulnerabilities. The attacks comprised DDoS, social engineering, and vulnerabilities in client-side online applications that allowed server-side systems to be compromised [7].

Failure to update vulnerable dependencies within deadlines can lead to pipeline breaks, exacerbating the effects of human errors [8].

Similarly, insecure build processes, facilitated by widely used tools such as Tekton, Jenkins, and GitHub Actions (GHA), create opportunities for intruders. The Apache Struts vulnerability, which exposed sensitive information from Equifax, exemplifies these vulnerabilities [9].

Addressing these challenges requires a multifaceted approach that combines technological innovation, best practices, and collaboration across organisational boundaries.

We aim to develop a system to identify unusual behaviour or deviations from expected patterns within the CI/CD pipeline and cloud platforms. This system intends to integrate into existing workflows to continuously monitor pipeline activities and cloud infrastructure.

The CI/CD pipeline makes software development more manageable, significantly automating code distribution and integration and accelerating the release cycle. However, there is always a constant risk associated with this flow. Security flaws may exist at every stage of the continuous integration and delivery (CI/CD) pipeline, from code commits to deployment.

Securing the CI/CD pipeline is critical since a breach anywhere along the pipeline can result in widespread implications such as unauthorised access, data loss, and code manipulation.

This study addresses the necessity of safeguarding the CI/CD pipeline, concentrating on potential risks and challenges and the usefulness of deep learning algorithms such as CNNs and LSTMs in detecting cyberattacks and forming log files of attack information.

Many previous studies recommended and implemented different types of CI/CD security measures, which are mainly based on static and dynamic application security testing (SAST and DAST) [10], source composition analysis (SCA) [11], access controls, and continuous monitoring and incident response [12]. However, these studies have drawbacks regarding attack detection rates, huge false positive rates, dependency and maintenance complexity, and resource intensive.

We aim to explore adaptive response mechanisms to mitigate detected anomalies or security threats. In this research, we employed two publicly available network traffic datasets named CSE-CIC-IDS2018 [13] and CSE-CIC-IDS2017 [13], which comprised different types of cyberattacks.

We are utilising these two datasets because they contain realistic and common cyberattack types, such as DDoS, brute force, botnets, etc. Both datasets include various features, such as flow, packet, and connection details, which support feature selection and analysis in machine learning models.

After that, we performed extensive data pre-processing techniques and performed optimal feature selection, which comprised extracting relevant features, data normalisation, and data resampling techniques.

Later, we leverage a hybrid deep learning (DL) algorithm, which is comprised of Convolutional Neural Networks (CNN) [14] and Long Short-Term Memory (LSTM) [15].

The trained CNN-LSTM model is deployed and loaded in the Jenkins pipeline. After the model integration, we monitor the CI/CD pipeline network activities and analyse the real-time data.

Finally, the model can identify potential cyberattacks and generate log files (Figures 6 and 7) in different CI/CD stages of the pipeline that resemble the types of attacks.

The main contributions of this work are given below:

1. We employed a hybrid CNN-LSTM model to detect different types of cyberattacks and utilized two network traffic datasets, CSE-CIC-IDS2018 and CSE-CIC-IDS2017, which resemble real-world traffic patterns.
2. We perform extensive data preprocessing techniques, including missing value handling missing values, and feature selection techniques comprised of feature selection using Random Feature Elimination (RFE) with Random Forests (RF), data normalisation, and data resampling techniques, Synthetic Minority Oversampling Technique (SMOTE) for oversampling and edited nearest neighbour (ENN) for undersampling.
3. We used the trained model, then deployed and integrated it in the CI/CD pipeline and continuously monitored the network traffic behaviour.
4. The model can predict seven types of cyberattacks inside the CI/CD pipeline and generate log files containing information about network anomalies.

The rest of the paper is as follows: Section 2 states the backgrounds and works of those related to our research, including detailed research gaps. Section 3 presents the methods of this study, where Research Objectives (Section 3.1) and Research Methodology (Section 3.2) are presented. Then, we report our results (Section 4) with working procedures and present our graphical output. Section 5 concludes our work with future works and the lessons learned from Sections 3 and 4.

## 2 Background and Related Work

CI/CD pipeline becomes indispensable for ensuring the efficiency and reliability of software development, integration, testing, and deployment processes.

Security issues persist throughout the entire application development and deployment lifecycle, including pre-and post-deployment phases in the cloud.

Despite their widespread adoption, several research gaps and challenges persist, hindering the realisation of their full potential.

Addressing the following gaps and challenges is essential for enhancing the effectiveness and scalability of the CI/CD pipeline in cloud environments. Findings from Mahboob and Coffman [16] and Huang, Minyan, et al. [17] support this.

Furthermore, numerous challenges arise across various security domains, including network and communication security [18], data privacy [16], and response time [19].

Moreover, security in every stage of the CI/CD pipeline is monitored by static application security testing (SAST), dynamic application security testing (DAST), and interactive application security testing (IAST) tools. These tools detect code vulnerabilities and identify and remediate security issues [10].

Source composition analysis (SCA) is used in CI/CD pipelines to detect open-source and license package vulnerabilities and ensure that software does not have insecure packages.

However, existing security measures have drawbacks in providing quick and real-time response, parallel pipeline jobs, flexible deployment and testing techniques, and generating vulnerability reports.

To address these challenges, a focus on AI-based anomaly detection emerges as a promising approach. This approach offers potential solutions for identifying and mitigating security threats in cloud environments.

Addressing the following gaps and challenges is indispensable for enhancing the effectiveness and scalability of the CI/CD pipeline in cloud environments by utilising AI techniques.

Research gaps include a limited focus on real-time ML system monitoring and a lack of emphasis on ML models' ethical and fair usage [20].

System observations and impact analysis for system modifications need more attention [21]. Also, the absence of proposals for efficient ML techniques in Software Engineering (SE) and future exploration of SE-ML fusion for scaling-up operations are notable research gaps [22].

The lack of discussion on real-world implementation challenges, limited focus on the scalability and adaptability of proposed frameworks, and the absence of detailed insights on framework integration with existing systems are critical areas requiring attention [23].

The limited research on bug report incompleteness in software maintenance and the effectiveness of follow-up questions in bug reports pose significant gaps in current studies [24].

The lack of exploration of kernel tracing impacts anomaly detection, and the potential for using other Natural Language Processing (NLP) techniques to enhance detection performance represents areas ripe for further investigation [25].

Similarly, the absence of specific examples of AI-driven CI/CD implementations, limited discussion of potential adoption challenges, and lack of comparative analysis with other emerging software development methodologies highlight important research aspects [26].

Lastly, the lack of focus on rigorous scientific methods in the Cybersecurity in Software Development and Systems (CSDS) domain and the need for developing best practices through experimentation and core research underscores the importance of advancing research efforts in this field [27].

Integrating machine learning (ML) technologies into various domains, including software engineering (SE), introduces new security challenges.

Dhabliya et al. (2024) emphasise the importance of securing ML ecosystems through robust monitoring and incident response mechanisms [20].

Despite this emphasis, there are gaps in the real-time monitoring of ML systems and ethical considerations in ML model usage. SafeOps, as proposed by Fayollas et al. (2020), introduces a framework for continuous safety assurance in autonomous vehicles (AVs) by integrating DevOps principles [21].

While SafeOps aims to improve safety and reliability, more focus should be placed on system observations and impact analysis for system modifications.

Abubakar et al. (2020) explore the interplay between ML and SE for software quality estimations [22]. While ML models show promise in accuracy, a gap exists in proposing efficient ML techniques in SE.

Future research could explore the fusion of SE and ML for scaling operations and enhancing tool integration.

The framework proposed by Aktas et al. (2023) integrates AI for anomaly detection in cloud computing systems, emphasising continuous monitoring and automation [23]. However, there is limited discussion on real-world implementation challenges and scalability issues.

Kohyarnejadfard et al. (2022) present an NLP-based approach for anomaly detection in microservice environments, achieving high accuracy [24]. Further investigation is needed to explore the impact of kernel tracing on anomaly detection and to leverage other NLP techniques for enhanced performance.

AI-driven CI/CD, as Mohammed et al. (2024) discussed, enhances software delivery processes, reducing manual effort and errors [25]. However, there is a lack of specific examples of AI-driven CI/CD implementations and discussions on potential challenges during adoption.

Chhillar and Sharma (2019) proposed Automated Continuous Testing (ACT) to enhance software reliability and speed up releases in cloud service models [26]. While ACT introduces innovative testing methodologies, future research could focus on improving software quality metrics and addressing security concerns.

As Lyubomir et al. (2021) discussed, evaluating machine learning models for IoT network security highlights the importance of robust algorithms like Support Vector Machine (SVM) and Random Forest [27]. However, challenges such as data limitations and resource-intensive classifiers need further exploration.

This synthesised section overviews the current landscape in securing ML ecosystems, continuous safety assurance, ML-SE interplay, anomaly detection, AI-driven CI/CD, automated continuous testing, and IoT network security. Further research in these areas can address existing gaps and contribute to advancing the field.

Unlike the research mentioned above gaps, our research objectives focused on enhancing the security of CI/CD pipelines in cloud environments through anomaly detection and anomaly log report generation.

To achieve this, we implemented deep learning (DL) based anomaly detection techniques, integrated them into the CI/CD pipeline workflow, and explored adaptive response mechanisms for mitigating security threats.

The CSE-CIC-IDS2018 and CSE-CIC-IDS2017 datasets were used to collect diverse network traffic data, including multiple attack types. Relevant features were extracted to train AI-based anomaly detection models. These models were then deployed into the CI/CD pipeline for real-time monitoring and threat detection. The system was evaluated for its effectiveness in improving security and reducing risks, focusing on detection accuracy and minimising false positives.

## 3 Methodology

This work proposes a deep learning (DL) based security architecture to advance software security and reliability in cloud environments. We provide insights, methodologies, and recommendations for enhancing the effectiveness and security of the CI/CD pipeline in cloud environments.

The proposed framework comprises six stages: data pre-processing, optimal feature selection, model learning, model deployment and integration, continuous monitoring, and finally anomaly detection.

Figure 1 shows the overall architecture of the proposed model for detecting network anomalies in the CI/CD pipeline.

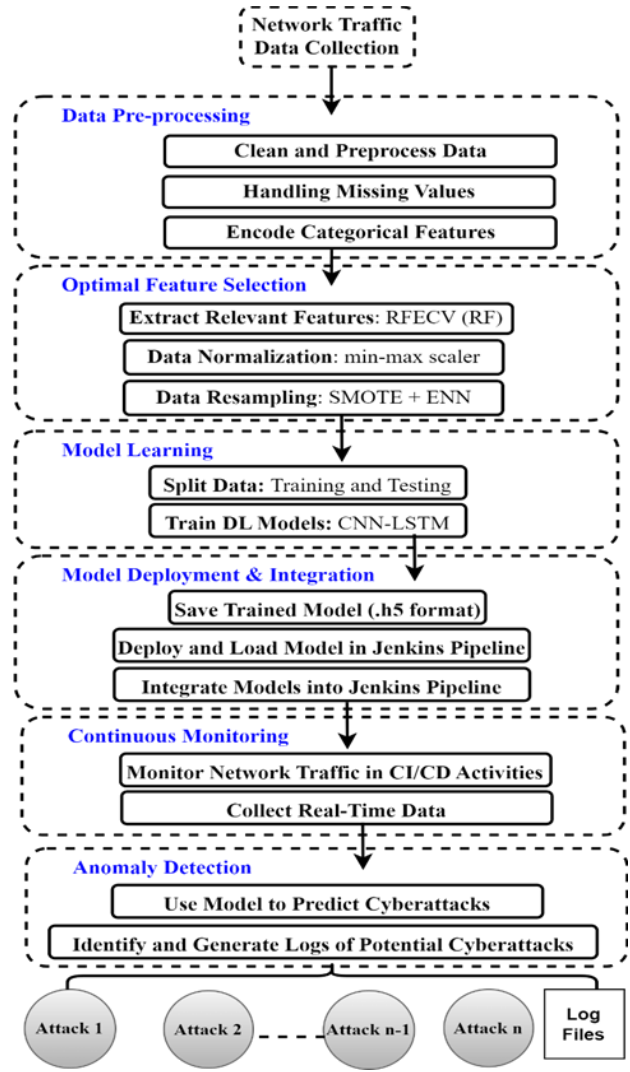

**Figure 1: Proposed Model Architecture**

### 3.1 Data Collection

Our research utilised the CSE-CIC-IDS 2018 and CSE-CIC-IDS 2017 [13] datasets published by the Canadian Institute of Cybersecurity (CIC).

CSE-CIC-IDS2018 [13] on AWS is a collaborative project aiming to generate benchmark datasets for intrusion detection. It features seven attack scenarios, including DoS, DDoS, bot, brute-force, Heartbleed, and web attacks, simulated on an AWS infrastructure, aligning with our objectives for enhancing CI/CD pipeline security through anomaly detection.

Utilising user profiles generates realistic benign traffic and describes attack scenarios. The dataset includes network captures, system logs, and 80 traffic features extracted using CICFlowMeter-V3, labelled based on attack schedules. This contribution aids in evaluating anomaly detection systems in real-world network environments.

The CSE-CIC-IDS2017 network traffic data consists of eight CSV files, each including five days of regular and irregular network activity from Monday to Friday. For testing, they set up an attack network with a router and switch and a victim network with a firewall, router, and switches.

Using CICFlowMeter software, 80 features were extracted from traffic-generated Pcap files, including regular and aberrant traffic. CIC-IDS2017 covers a range of assaults, including portscan, infiltration, brute force, DDoS, web, and botnet.

### 3.2 Data Pre-processing

The datasets we utilised are both generated in a simulated environment that depicts the real-world traffic flow and collected 80 different network traffic features. However, we excluded various features that resulted in abnormal traffic behaviour.

We also calculated the percentage of zeros in each feature and removed a feature if more than 30% of samples were zero. The data samples with missing values are removed from the data frame.

We also removed time-related and switch-related features. We used a label encoder to encode categorical features like 'protocol' and 'service'. Web attacks are grouped because they exhibit similar network traffic characteristics to brute force, XSS, and SQL injection.

Additionally, several DoS assaults, such as DoS slowHTTP, DoS hulk, DoS slowloris, and DoS goldeneye, are classified as DoS.

### 3.3 Optimal Feature Selection

Feature selection is selecting the most relevant features from an extensive set and removing irrelevant ones to enhance the model performance and reduce computational loads on prediction models.

Removing superfluous and noisy features improves learning efficiency and reduces model training time. Furthermore, feature significance assesses the value of each feature in the prediction model.

After pre-processing the data, we perform feature selection techniques to identify the relevant features from both datasets. We employed Random Feature Elimination (RFE) with Random Forest (RF) to select the relevant features. RFE used wrapper-style and filter-based feature selection techniques, initially searching a subset of features from the entire dataset and removing them until the desired number of features remained. RFE with RF iteratively removed all the less important features from the datasets with the least contribution to the predictive models.

RFE commences by training a Random Forest model on the entire dataset. The Random Forest algorithm determines how important a feature is based on how much a feature reduces prediction errors. The least significant characteristics are eliminated in each iteration, and the model is then retrained using the remaining features. This process is repeated until a performance condition is satisfied or the required number of features is reached.

After the feature selection, we normalise our entire data to scale our dataset between 0 and 1. It is a common approach in machine learning (ML) because we do not know about all the data points in the entire dataset. By transforming it between 0 and 1, we ensure that all the data points are contributed equally to the model. Normalising data from multiple scales ensures equitable contribution during model training. Scaling the data equalises all features, allowing for faster convergence and optimisation using gradient descent. The mathematical formulation for the min-max scaler is given below:

$$\min - \max = \frac{x - \min(x)}{\max(x) - \min(x)} \quad (1)$$

After data normalisation was achieved, we performed data resampling techniques, where we used the Synthetic Minority Oversampling Technique (SMOTE) [28] to oversample the minority class and edited nearest neighbour (ENN) to undersample the majority class.

In ML, imbalanced data is a classification problem where all the classes are not distributed equally in the entire dataset.

As a result, imbalanced data is biased towards one or more classes, with few samples for others. Training a model with uneven data can lead to bias toward one or two classes. To overcome this problem, we utilised SMOTE and ENN. SMOTE uses Euclidean distance to calculate sample distances, followed by modification of the k-nearest neighbour. After taking n number of samples, they calculate the imbalance ratio and determine the required samples. The created samples (y) are taken from the k-nearest neighbour and used to create fresh synthetic samples. The majority class was chosen as an undersampling strategy. ENN removes the majority of the k-nearest neighbours sample.

### 3.4 Model Learning

After completing the feature selection steps, we split our datasets into training and testing data. Training data is used to train the deep learning (DL) model, and testing is used to evaluate the model's performance in unseen data. Overall, 80% of the data are used for model training, and the remaining 20% are used to evaluate the predictive model's performance.

Our experiment used a combination of CNN [14] and LSTM [15]; CNN-LSTM. CNNs are a type of deep neural network that is commonly used to analyse visual images. They include convolutional layers, pooling layers, and fully linked layers. Convolutional layers use filters on input data to detect different features. Pooling layers minimise the spatial dimensions of the data, making the representations more manageable and less prone to distortion and translation.

The identified features are integrated into fully linked layers at the network's end to create final predictions. LSTMs are a form of recurrent neural network (RNN) that models sequential data and detects long-term dependencies. They have gates (input, output, and forget gates) that control the flow of information, allowing the network to learn which information to keep or reject. This architecture enables LSTMs to recall essential information over extended periods, making them useful for sequence prediction problems.

The CNN-LSTM model combines the capabilities of CNNs and LSTMs, making it ideal for applications that require both spatial and temporal data. The architectural Overview of CNN-LSTM is given below:

*CNN Component:* The earliest layers include convolutional and pooling layers that extract spatial features from the input data. These traits may signify key local trends or attributes. We used three convolutional and pooling layers followed by two dropout layers. The rectified linear activation function (ReLU) is used as the activation function.

*Flattening Layer:* Following the convolutional layers, the data is usually flattened into a one-dimensional vector containing all retrieved features.

*LSTM Component:* The flattened vector is fed into LSTM layers, which examine the temporal correlations and dependencies among the retrieved features. We used two LSTM cells in the experiments.

*Fully Connected Layer:* The final prediction is produced by passing the final output of the LSTM layers through fully connected layers. A SoftMax activation function is used in this layer.

Overall, we used Adam as the optimiser with a learning rate of 0.001, categorical cross-entropy to measure the loss, and the weighted average accuracy of the model.

### 3.5 Model Deployment & Integration

After training the model in the model training phase, we saved the trained model in a hierarchical data format (.h5).

We load the model in the Jenkins server for CI/CD pipeline integration. For our research, we employed a plugin titled "Machine Learning" [23], provided by Jenkins, to seamlessly integrate our model within the CI/CD Pipeline.

This plugin integrates Machine Learning workflows, including Data preprocessing, Model Training, Evaluation, and Prediction, with Jenkins build tasks.

This plugin can execute code fragments via the IPython kernel currently supported by Jupyter. We integrate the CNN-LSTM model into the Jenkins pipeline. Once the model has been trained and validated, it must be packaged as a deployable artefact.

This includes saving the model weights, configurations, and any dependencies needed to execute the model. Jenkins automatically runs tests and validation processes whenever new code is posted to the repository, ensuring that code changes are continuously integrated. This helps to identify bugs early on and guarantees that the codebase remains stable.

### 3.6 Continuous Monitoring

Jenkins offers real-time monitoring of pipeline executions. It tracks the status of each stage, making it simple to detect and debug errors. It also checks the deployed model's performance by measuring error rates, resource utilisation, and response time.

We use automated scripts in the Jenkins workflow to identify model predictions or behaviour anomalies. For example, if the model detects an attack, it will record the information in the log file. Jenkins can also help retrain and redeploy the pipeline with new data.

### 3.7 Evaluation Criteria

The unified data frame was partitioned 80 to 20 for training and testing purposes. Performance indicators, including accuracy, precision, recall, and f1 score, were used to evaluate model performance and optimise hyperparameters.

We used a confusion matrix to visualise classification performance and output in matrix format. A confusion matrix categorises classification model outcomes as true positive (TP), true negative (TN), false positive (FP), or false negative (FN). A confusion matrix can help describe the outcomes of a classification model. The definitions are as follows:

$$Accuarcy = \frac{TP+TN}{TP+TN+FP+FN} \quad (2)$$

$$Precision = \frac{TP}{TP+FP} \quad (3)$$

$$Recall = \frac{TP}{TP+FN} \quad (4)$$

$$F1\ Score = 2\frac{Precision*Recall}{Precision+Recall} \quad (5)$$

### 3.8 Software and Hardware Requirements

In this experiment, a Windows 10 PC with a 4 GB NVIDIA GeForce RTX 3050 graphics processing unit (GPU), an AMD Ryzen 9 5900HX processor, 16 GB of RAM, and a 512 GB solid-state drive (SSD) was used.

Several Python 3.7 modules, such as Keras [29], TensorFlow 2.8.0 [30], and Scikit-learn [31], were used to create the Network Intrusion Detection System (NIDS) model.

In addition, NumPy was used for numerical calculations, the Pandas library was used for data analysis, and Matplotlib and Seaborn were used to create graphical representations of the experimental results.

## 4 Results and Discussions

In our experiment, we performed feature selection using RFE (RF), through which we selected the optimal number of features that contribute to predicting the different types of network anomalies. Figures 2 and 3 show the selected features from both datasets.

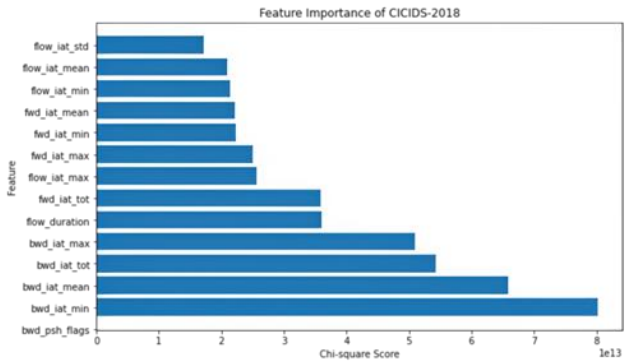

**Figure 2: Feature Importance of CSE-CICIDS2018**

After the feature selection, we applied data resampling techniques for both datasets, as they were imbalanced. Table 1 shows the data distribution before and after the data resampling.

Our experiments used a CNN-LSTM architecture to detect the CI/CD pipeline network anomalies. Table 2 shows the proposed CNN-LSTM's performance comparison with other state-of-the-art experiments in the CSE-CICIDS2017 dataset.

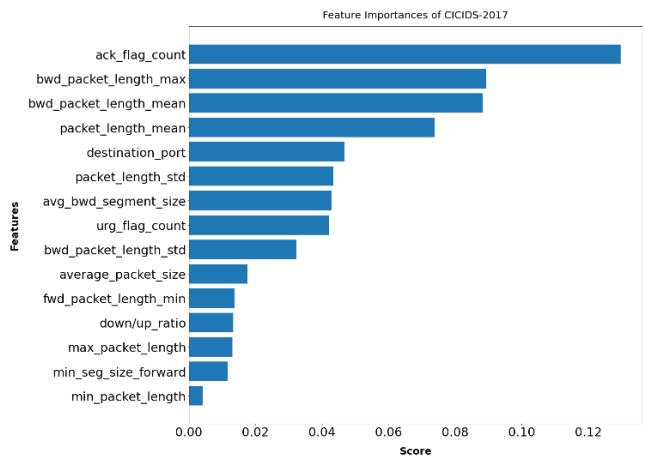

**Figure 3: Feature Importance of CSE-CIC-IDS2017**

The proposed CNN-LSTM outperforms the existing works with a high accuracy rate of 98.30%. Figure 4 shows the confusion matrix of the proposed CNN-LSTM model for the CSE-CICIDS2017 dataset.

**Table 1: Data Resampling of the Datasets**

| Anomalies | CIC-IDS2107 | | CIC-IDS2018 | |
|---|---|---|---|---|
| | Before Resampling (%) | After Resampling (%) | Before Resampling (%) | After Resampling (%) |
| Benign | 26 | 20 | 33 | 21 |
| DoS | 24 | 17 | 19 | 14 |
| DDoS | 22 | 16 | 13 | 15 |
| Web | 19 | 17 | 1 | 9 |
| Portscan | 7 | 11 | 15 | 14 |
| Bot | 1 | 9 | - | - |
| Brute Force | 1 | 10 | 19 | 13 |
| Infiltration | - | - | 9 | 14 |

The proposed model can detect attacks with more than 97% accuracy, except for Bots, which can be detected only 88.3% of the time. The rest are misclassified by benign network traffic.

**Table 2: Comparison of CNN-LSTM for CIC-IDS2017 with Existing Work**

| Model | Accuracy | Precision | Recall | F1 Score |
|---|---|---|---|---|
| CNN [32] | 97.5 | 97.7 | 97.6 | 97.3 |
| RBF-BLS [33] | 96.63 | - | 96.87 | - |
| Bi-RNN GRU [34] | 98.99 | - | - | - |
| **CNN-LSTM (proposed)** | **98.30** | **98.45** | **98.30** | **98.34** |

For CSE-CICIDS 2018, our model achieves higher accuracy than existing works. It receives an accuracy of 98.69%, higher than other recent works.

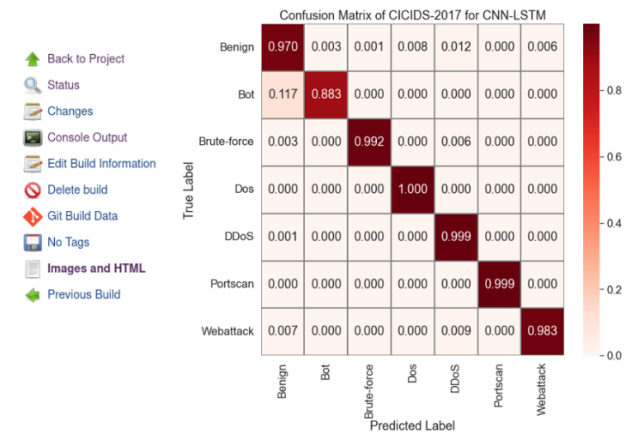

**Figure 4: Confusion Matrix of CSE-CIC-IDS 2017**

It performs better because merging CNN with LSTM allows us to take advantage of CNN's effectiveness in extracting features from unprocessed data and LSTM's comprehension of temporal connections.

Figure 5 shows the confusion matrix for the CSE-CIC-IDS 2018 dataset. It can detect almost all types of attacks, more than 96%. It can detect 95.7% of infiltration and 89% of web attacks, whereas 8.1% were misclassified as brute-force attacks.

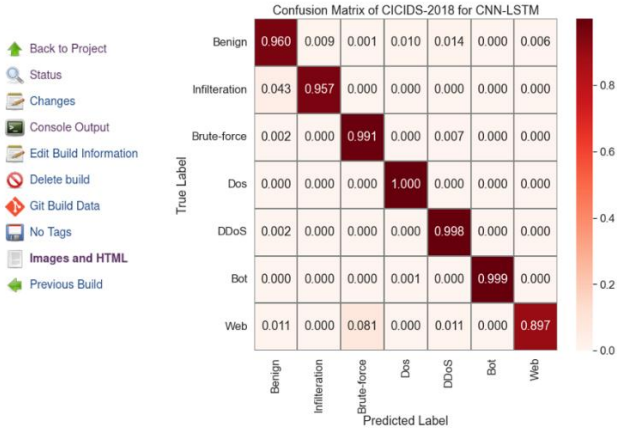

Figure 5: Confusion Matrix of CSE-CIC-IDS 2018

Figure 6 shows the log file output while monitoring the CI/CD pipeline. This log report gives the traffic details, types of network anomalies, and timestamps in the monitored network traffic. We also generated the log report while monitoring the Jenkins CI/CD pipeline, where log data shows information about traffic flows, attack types, timestamps, and types of network traffic.

Table 3: Comparison of CNN-LSTM for CIC-IDS2018 with Existing Work

| Model | Accuracy | Precision | Recall | F1 Score |
|---|---|---|---|---|
| LSTM + AM [35] | 96.2 | - | - | - |
| Light GBM [36] | 97.5 | - | - | - |
| CFBLS + BLS [33] | 97.46 | - | - | 97.45 |
| **CNN-LSTM (proposed)** | **98.69** | **98.73** | **98.68** | **98.70** |

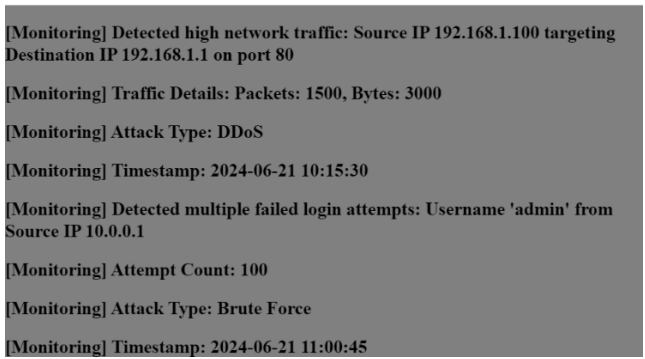

Figure 6: Log files of Jenkins Pipeline (Monitoring Stage)

We also captured the log file during different CI/CD pipeline stages, such as the build, test, deploy, and monitoring stages. Figure 7 shows the sample output for different stages of the Jenkins pipeline, the build, test, deploy, and monitoring stages.

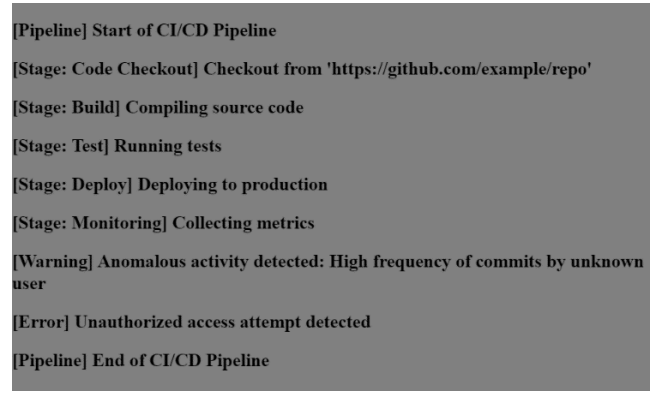

Figure 7: Log files of Jenkins Pipeline (Different Stages)

## 5   Conclusions and Future Works

This study has demonstrated the efficacy of utilising AI-based anomaly detection techniques, particularly Convolutional Neural Network (CNN) and Long Short-Term Memory (LSTM) architectures, to enhance the security and reliability of the CI/CD pipeline in cloud platforms.

We utilised two popular network traffic datasets and performed extensive data pre-processing, and applied data resampling techniques using SMOTEENN after selecting the relevant features using RFE (RF).

Our proposed CNN-LSTM outperforms the existing works by achieving an accuracy of 98.69% and 98.30% in the CSE-CICIDS2018 and CSE-CICIDS2017 datasets, respectively.

By integrating these techniques into the CI/CD pipeline workflow and exploring adaptive response mechanisms, we have identified unusual behaviour or deviations from expected patterns within CI/CD pipeline activities and cloud infrastructure.

The results indicate that the implemented anomaly detection system contributes significantly to mitigating security threats and enhancing the overall integrity of software delivery processes.

Through rigorous experimentation and evaluation, we have shown this approach's practical feasibility and effectiveness in addressing security challenges in modern DevOps practices.

The main limitations of the proposed study are that our attack surface is fixed, and the system is unable to detect new network anomalies. Moreover, our system was not tested in a real-world software development project, and we did not utilise any existing cloud infrastructure.

In the future, the scalability and applicability of the AI-based anomaly detection system can be extended to handle larger and more complex CI/CD pipeline environments and diverse cloud platforms like AWS, Google Cloud, and Microsoft Azure.

Additionally, incorporating real-time monitoring capabilities and adaptive learning mechanisms could enhance the system's responsiveness to emerging security threats.

Furthermore, investigating the integration of other AI techniques, such as reinforcement learning and ensemble methods, may offer additional insights into optimising anomaly detection performance.

Overall, future work in this area should focus on enhancing AI-driven anomaly detection systems' robustness, scalability, and effectiveness for safeguarding the CI/CD pipeline and cloud environments against emerging security threats.